\documentclass[pra,twocolumn,a4paper,showpacs,superscriptaddress]{revtex4}
\usepackage{amsmath}
\usepackage{amsfonts}
\usepackage{graphicx}

\begin{document}

\title{Separability of a family of one parameter W and  GHZ multiqubit states 
using Abe-Rajagopal $q$-conditional entropy approach}
\author{R. Prabhu}
\affiliation{Department of Physics, Kuvempu University, 
Shankaraghatta, Shimoga-577 451, India}
\author{A. R. Usha Devi}
\email{arutth@rediffmail.com}
\affiliation{Department of Physics, Bangalore University, 
Bangalore-560 056, India}
\affiliation{Inspire Institute Inc., McLean, VA 22101, USA.} 
\author{G. Padmanabha}
\affiliation{Department of Physics, Bangalore University, 
Bangalore-560 056, India}

\date{\today}

\begin{abstract}
We employ conditional Tsallis $q$ entropies to study the 
separability of symmetric one parameter W and GHZ multiqubit mixed 
states. The strongest limitation on  separability is realized 
in the limit $q\rightarrow\infty$, and is found to be 
much superior to the condition obtained 
using the von Neumann conditional entropy (q=1 case). 
Except for the example of  two qubit 
and  three qubit symmetric states of  GHZ family, the $q$-conditional 
entropy method leads to sufficient - but not necessary - conditions on 
separability.
\end{abstract}

\pacs{03.67.Mn, 03.65.Ud}
\maketitle

\section{INTRODUCTION}

Quantum entanglement has evoked intense interest in recent years as it occupies a 
central position in quantum computation and information theory~\cite{Zoll,Horreview}. 
Characterizing whether a given composite quantum state is separable or 
entangled is a key issue in this currently emerging field. 
Entropic characterization~\cite{entropy,Horodecki,AR1,Tsallis,Abe,Ros,Batle1,Batle2} proves to be 
significant  
in this direction. One of the important observations is that the subsystems of 
a separable state $\hat\rho^{(\rm sep)}_{AB}=\sum_i\, p_i\, 
\hat\rho_A^{(i)}\otimes\hat\rho^{(i)}_{B}$ 
(with  $0\leq p_i\leq 1,\ \sum_i\, p_i=1$), are more {\em ordered} than the whole system i.e., 
$S(\hat\rho^{(\rm sep)}_{AB})\geq S(\hat\rho_A),\, S(\hat{\rho}_B)$,  where
$ S(\hat\rho)=-{\rm Tr}[\hat\rho\log\hat\rho]$ denotes the von Neumann entropy. 
In contrast, an arbitrary pure entangled state  satisfies the inequality 
\begin{equation}
\label{cond1}
S(B|A)=S(\hat\rho_{AB})-S(\hat\rho_A)\leq 0,  
\end{equation}      
reflecting the remarkable fact~\cite{note} that {\em  pure entangled states 
are more disordered locally than globally.}  
Negative conditional entropies (implied by the inequality~(\ref{cond1})) 
provide sufficient - but not necessary -  criterion to characterize mixed entangled states. 
In the case of two qubit  Werner state,  $\hat\rho_{AB}=|\psi _{AB}\rangle\langle\psi 
_{AB}|\,x+I_4\,(1-x)/4;$\,
${\scriptsize 0\leq x\leq 1,\, |\psi _{AB}\rangle=\frac{1}{2}\,
(|\uparrow _{A}\,\uparrow _{B}\rangle+|\downarrow _{A}\,\downarrow _{B}\rangle)}$   
the conditional entropic criterion Eq.~(\ref{cond1}) leads to $0~\leq~x~\leq~0.747$ as the range 
of separability ( the von Neumann conditional entropy 
is positive in this range of the parameter $x$), which is clearly weaker compared to that 
obtained through Peres' partial transpose criterion~\cite{Peres}: $0~\leq~x~\leq~\frac{1}{3}.$ 
This example brings out the limitation of the  entropic inequality (\ref{cond1}), 
in characterizing entanglement in mixed composite states.  Generalized entropic 
measures~\cite{Horodecki,AR1,Tsallis,Abe,Ros,Batle1,Batle2}
provide more sophisticated tools to explore global vs local disorder in 
mixed states and lead to more stringent limitation on separability than that obtained using
positivity of the conditional von Neumann entropy. 
In this context, the quantum counterparts of the R\'{e}nyi entropy~\cite{Horodecki}, 
$S_q^{(R)}(\hat\rho)=\frac{1}{1-q}\, \log {\rm Tr}\,[\hat\rho^q],$ and the 
Tsallis entropy~\cite{Tsallis2}, $S_q^{(T)}(\hat\rho)=\frac{{\rm Tr}\,[\hat\rho^q]-1}{1-q}$ 
have often been employed. In the limit $q\rightarrow 1$ both these generalized entropic 
measures~\cite{note2} reduce to the von Neumann entropy. 
 Horodecki et. al.~\cite{Horodecki} recognized that 
$S_q^{(R)}(\hat\rho^{\rm (sep)}_{AB})\geq S_q^{(R)}(\hat\rho_A),S_q^{(R)}(\hat\rho_B)$ 
for separable states and thus negative values of the conditional R\'{e}nyi entropy 
$S_q^{(R)}(B|A)=S_q^{(R)}(\hat\rho_{AB})-S_q^{(R)}(\hat\rho_{A})$ (with 
 $S_q^{(R)}(\hat\rho_{A})$ being the maximum of the subsystem R\'{e}nyi entropies) 
is a signature of quantum entanglement. On the other hand, based on Tsallis entropy and 
the form invariant structures of Khinchin's axioms, Abe and Rajagopal~\cite{AR1}
generalized the concept of conditional entropy as 
\begin{eqnarray}
\label{AR}
S^{(T)}_q(B|A)&=&\frac{S^{(T)}_q(\hat\rho_{AB})-S^{(T)}_q(A)}{1+(1-q)\,S^{(T)}_q(\hat\rho_A)}
\nonumber\\
        &=&\frac{1}{q-1}\left[1-\frac{{\rm Tr}\,[\hat\rho^q_{AB}]}
		{{\rm Tr}\,[\hat\rho^q_{A}]}\right],
\end{eqnarray} 
where $S^{(T)}_q(\hat\rho_{AB}),\,S^{(T)}_q(\hat\rho_A),$ respectively denote the Tsallis 
$q$-entropies associated with $\hat\rho _{AB}$ and its subsystem density 
operator $\hat\rho_A$. 
The Abe-Rajagopal (AR) $q$-conditional entropy, given by 
Eq.~(\ref{AR}), is  nonnegative for a separable 
state, but may assume negative value~\cite{note3} in a 
quantum entangled state, suggesting its importance in   
the characterization of quantum entanglement in mixed 
composite states.  In the particular case of two qubit Werner state, 
Abe and Rajagopal~\cite{AR1} recovered  the necessary and 
sufficient condition for separability 
(i. e., $0~\leq~x~\leq~\frac{1}{3}$), by examining the positivity of 
$q$-conditional entropy in the limit $q\rightarrow \infty.$
Employing the same approach,  Tsallis et. al.,~\cite{Tsallis} re-discovered the Peres criterion 
  for separability in  more general two qubit mixed states. 
As a further extension,  Abe~\cite{Abe} showed that the negativity 
of $q$-conditional entropy gives the correct range of inseparability for generalized 
Werner states of $N$-qudits.  Batle et. al.~\cite{Batle1,Batle2} performed a comprehensive 
numerical survey of the space of bipartite systems and identified that the volume 
occupied by the states with positive conditional entropies $S^{(R)}_q(B|A)$ and $S^{(T)}_q(B|A)$ 
decreases monotonically as $q$ increases (it is found that this monotonic behavior  
is more pronounced when  Tsallis $q$-conditional entropies are investigated~\cite{Batle2}). 
Moreover, the numerical investigation~\cite{Batle2} indicated that the Peres criterion provides 
a  much stronger condition on separability than that derived from the 
positive conditional entropies in the  $q\rightarrow\infty$ limit.  

Positivity of  conditional entropies, $S^{(R,T)}_q(B|A),$ is based entirely on 
the  global and local spectra of the composite state and is basically, 
one of the implications of majorization~\cite{Niel, Wolf}, which is 
the strongest spectral criterion of separability.  It has been shown~\cite{Niel} 
that any criteria, based only on the eigenvalues of the state and 
its reductions, do not provide a complete characterization of entanglement. 
There are examples of entangled states, which can not be detected by 
any spectral criteria, as separable states with the same global and local 
spectra exist~\cite{Niel,Wolf}. 
So, the $q$-conditional entropic characterization does not lead, in 
general, to  necessary and sufficient condition for separability. However, 
this method is fruitful in obtaining more stringent limitations on separability, 
than the one identified from the familiar $q=1$  case (von Neumann conditional entropy). Further, 
negative $q$-conditional entropy implies 
that the composite state is distillable,   as it signals violation of the reduction 
criterion~\cite{Wolf}. 

In the present paper we investigate separability of one parameter symmetric 
multiqubit W and GHZ states using the AR $q$-conditional entropy 
approach. The strongest limitation on separability, 
is obtained in the $q\rightarrow\infty$ limit, 
and is found to agree with the Peres' criterion 
only for two and three qubit states of the GHZ family. We obtain the explicit 
$N$ dependent range of separability based on the positivity of  
$q$-conditional entropies, $S^{(T)}(B|A),$ in the limit $q\rightarrow\infty$.  
 In Sec.~II, we discuss $q$-entropic characterization 
of separability in  two and three qubit 
symmetric states and compare the results with that obtained from the
Peres criterion. In Sec.~III we derive the 
strongest constraints (obtained in the limit $q\rightarrow \infty$) 
on separability for the one parameter family of W and GHZ multiqubit states. 
A summary of results is given in Sec.~IV.
 
\section{ Two and three qubit symmetric states}
\label{sec:Two}

Symmetric states of qubits~\cite{stockton} are those, which remain unaltered under 
permutations~\cite{footnote} of the qubits. They have attracted 
 a great deal of attention recently~\cite{ARU} due to mathematical elegance 
 offered in characterizing them 
 as well as  for their experimental significance~\cite{Expt}.
Symmetric multiqubit states get restricted - due 
to permutation symmetry - to a $N+1$ dimensional subspace 
$\{|\frac{N}{2},M\rangle;\,-\frac{N}{2}\leq M\leq\frac{N}{2}\}$ of the 
entire $2^N$ dimensional Hilbert space ${\cal H}=(C^2)^{\otimes N}$ of 
$N$ qubits. (Here $|\frac{N}{2},M\rangle$ denote the simultaneous eigenstates 
of the squared total angular momentum operator $\hat J^2$ and the  $z$ component 
$\hat J_z$, with $\hat {\vec J}=\frac{1}{2}\sum _{\alpha}\hat{\vec\sigma} _{\alpha};
\,\alpha=1,2,\ldots,N;\,\,\hat{\vec\sigma} _{\alpha}$ being the Pauli operator of the 
$\alpha^{th}$ qubit). 

In the following, we will be focusing on the AR $q$-conditional entropy 
characterization of a simple one parameter mixed state of two qubits given 
by
\begin{equation}
\label{wstate2}
\hat\rho_{AB}=\left(\frac{1-x}{3}\right)\,P_{N=2}+x\,\,|\Phi^+_{AB}\rangle
\langle \Phi^+_{AB}|;\,\,0\leq x\leq 1
\end{equation}
where $P_2=\sum _{M}\,\,\,|1,\,M\rangle\langle 1,\,M|;\,M=-1,0,1,$ 
corresponds to projection operator onto the symmetric subspaces of 
two qubit states characterized by the maximum value of total angular momentum 
$J=N/2=1$. In terms of the standard two qubit basis, the symmetric states 
$\{|1,M\rangle \}$ are given by 
$|1,1\rangle=|\uparrow _A\uparrow _B\rangle,$ $|1,-1\rangle~=~
|\downarrow _A\downarrow _B\rangle$  and $|1,\,0\rangle~=~
\frac{1}{\sqrt 2}\,\,[|\uparrow _A\,\downarrow _B\rangle
+|\downarrow _A\,\uparrow _B\rangle]=|\Phi^+_{AB}\rangle$, one of the
 Bell states.

It is easy to identify the eigenvalues of $\hat\rho _{AB}$;
\begin{eqnarray}
\label{2eigen}
p_1(AB)&=&p_2(AB)=\frac{1-x}{3},\nonumber\\
p_3(AB)&=&\frac{1+2x}{3},\,\,\,\,p_4(AB)=0.
\end{eqnarray}
The reduced subsystems density matrices of $\hat\rho _{AB}$ are given by
\begin{eqnarray}
\label{redrho}
\hat\rho _A&=&{\rm Tr}_B\,[\hat\rho _{AB}]=\frac{1}{2}\,I_A\nonumber\\
\hat\rho _B&=&{\rm Tr}_A\,[\hat\rho _{AB}]=\frac{1}{2}\,I_B,
\end{eqnarray} 
where $I_{A(B)}=|\uparrow _{A(B)}\rangle\langle\uparrow _{A(B)}|
+|\downarrow _{A(B)}\rangle\langle\downarrow _{A(B)}|$~
denotes the identity operator in the subspace of individual qubits. Thus 
the eigenvalues of $\hat\rho_A$ ($\hat\rho_B$) are $p_k(A\ {\rm or}\ B)~=~1/2;\ k=1,2.$  
 
\begin{figure}[h]
 \includegraphics*[width=2.15in,keepaspectratio]{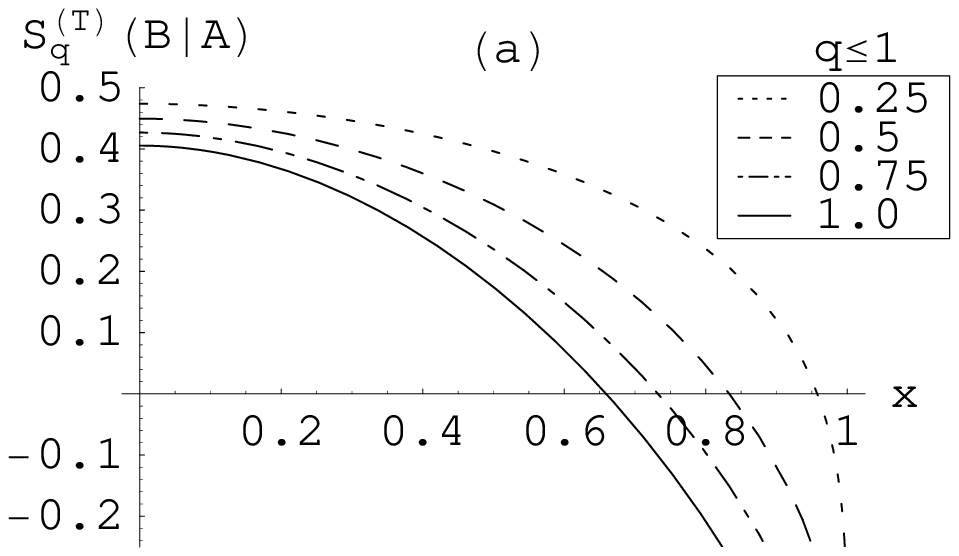} 
 \includegraphics*[width=2.15in,keepaspectratio]{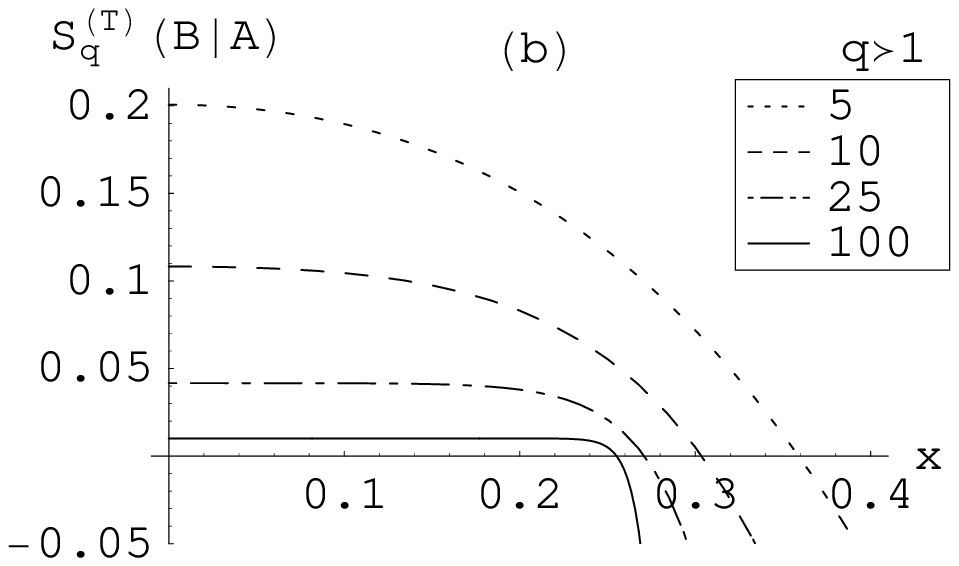}
 \caption{The AR $q$-conditional entropy $S^{(T)}_q(B|A)$ with (a) $q\leq 1$ 
 and (b) $q>1$ for the two qubit mixed state $\hat\rho _{AB}$ of 
 Eq.~(\ref{wstate2}) as a function of the parameter $x$. Note that 
 $\lim_{q\rightarrow 1}S^{(T)}_q(B|A)=0$, i.e., vanishing von Neumann conditional entropy, 
  leads to $x=0.6593,$ which is clearly lower than the corresponding 
 value obtained by solving the equation $S^{(T)}_q(B|A)=0$ with $q>1$ justifying the 
 superiority of the $q$-conditional entropy approach. All quantities are dimensionless.}
  \label{fig:one}
\end{figure}

\begin{figure}
\includegraphics*[width=2.15in,keepaspectratio]{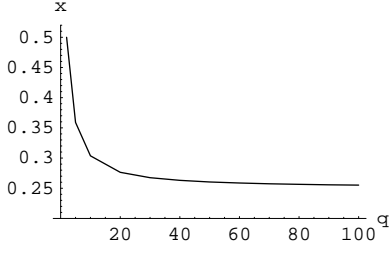}
 \caption{Implicit plot of $S^{(T)}_q(B|A)=0$ as a function of  $q$, 
 for the two qubit state $\hat\rho _{AB}$ of Eq. (\ref{wstate2}). In the limit 
 $q\rightarrow\infty$ it may be seen that $x\rightarrow 0.25$. All quantities are 
 dimensionless.}
  \label{fig:two}
\end{figure}

We now construct the AR $q$-conditional entropy (see Eq.~(\ref{AR}) for definition) for 
the two qubit symmetric state of Eq.~(\ref{wstate2}) as,
\begin{widetext}
\begin{eqnarray}
S^{(T)}_q(B|A)=\frac{1}{q-1}\left[1-\frac{\displaystyle{\sum _{i=1}^4}\, [p_i(AB)]^q}
{\displaystyle{\sum _{k=1}^2}\, [p_k(A)]^q}\right]
=\frac{1}{q-1}\left[1-\frac{(\frac{1-x}{3})^q
+\frac{1}{2}(\frac{1+2x}{3})^q}{(\frac{1}{2})^q}\right].
\end{eqnarray}
\end{widetext}
Non-positive value of $S^{(T)}_q(B|A)$ necessarily implies that the state 
$\hat\rho _{AB}$ of Eq.~(\ref{wstate2}) is quantum entangled. To find 
the minimum value of $x$ above (below) which the $q$-conditional entropy 
$S^{(T)}_q(A|B)$ takes negative (positive) values, we plot $S^{(T)}_q(B|A)$ as a 
function of $x$ for different choices of the parameter $q$ in Fig.~(\ref{fig:one}). It is 
evident that the value of $x$ for which $S^{(T)}_q(B|A)\rightarrow 0$ keeps 
reducing with the increase of $q$. So, the strongest constraint on separability 
is obtained in the limit $q\rightarrow\infty$. An implicit plot of 
$S^{(T)}_q(B|A)=0$ (see Fig.~\ref{fig:two}) clearly indicates $x\rightarrow\frac{1}{4}$ 
as $q\rightarrow\infty$. It is illuminating to note that the Peres 
Horodecki criterion also gives $x\leq\frac{1}{4}$ as  necessary 
and sufficient condition for separability of the two qubit
state $\hat\rho _{AB}$ of Eq.~(\ref{wstate2}). 

We now consider a three qubit generalization of the example considered in 
Eq.~(\ref{wstate2}): 
\begin{equation}
\label{three}
\hat\rho_{ABC}=\left(\frac{1-x}{4}\right)P_{N=3}+x\,|W\rangle\langle W|
\end{equation}
where 
\begin{eqnarray}
|W\rangle&=&\frac{1}{\sqrt 3}\,
(|\downarrow_A \uparrow_B\uparrow_C\rangle+|\uparrow_A\downarrow_B \uparrow_C\rangle+
|\downarrow_A \downarrow_B\uparrow_C\rangle)\\
&=&\left|\frac{3}{2},\,\frac{1}{2}\right\rangle.\nonumber
\end{eqnarray}
The projection operator is given by 
${\scriptsize P_3=\sum _{M}|3/2,M\rangle\,\langle 3/2,M|;\, M~=~3/2,\,1/2,-1/2,-3/2,}$ and the 
symmetric states  $\{|3/2,\,M\rangle\}$ are given 
 explicitly  in terms of the three qubit basis states as follows:
 {\scriptsize
\begin{eqnarray*}
\left|\frac{3}{2},\,\frac{3}{2}\right\rangle&=&
|\uparrow _A\,\uparrow _B\,\uparrow _c\rangle\\
\left|\frac{3}{2},\,\frac{1}{2}\right\rangle&=&|W\rangle
     =\frac{1}{\sqrt 3}\,
     (|\downarrow_A \uparrow_B\uparrow_C\rangle+|\uparrow_A\downarrow_B \uparrow_C\rangle
     +|\downarrow_A \downarrow_B\uparrow_C\rangle)\\
\left|\frac{3}{2},-\frac{1}{2}\right\rangle&=&
     \frac{1}{\sqrt 3}\,(|\uparrow _A\,\downarrow _B\,\downarrow _c\rangle
	 +|\downarrow _A\,\uparrow _B\,\downarrow _c\rangle
	 +|\downarrow _A\,\downarrow _B\,\uparrow _c\rangle)\\
\left|\frac{3}{2},-\frac{3}{2}\right\rangle&=&|
     \downarrow _A\,\downarrow _B\,\downarrow _c\rangle.\\
\end{eqnarray*}}

Expressing $\hat\rho _{ABC}$ in the total angular momentum basis states 
$\{|3/2,M\rangle,\,M=3/2,\,1/2,-1/2,-3/2\},$  we obtain 
\begin{equation}
\hat\rho_{ABC}=\left(
\begin{array}{cccc}
\frac{1-x}{4} & 0 & 0 & 0 \\
0 & \frac{1+3x}{4} & 0 & 0 \\
0 & 0 & \frac{1-x}{4} & 0 \\
0 & 0 & 0 & \frac{1-x}{4} \\
\end{array}\right)
\end{equation}
from which  the nonzero eigenvalues of the three qubit state 
$\hat\rho _{ABC}$ are easily found to be 
\begin{equation}
\label{w3ev}
\frac{1+3x}{4},\,\,\frac{1-x}{4}\,\,({\rm three\,times}).
\end{equation}
The two qubit marginal density matrix $\hat\rho_{AB}$ is given by~\cite{footnote2}
\begin{equation}
\label{two}
\hat\rho _{AB}=\left(\frac{1-x}{3}\right)P_2+\frac{x}{3}[|1,1\rangle\langle 1,1|
                 +2|\Phi ^+_{AB}\rangle\langle\Phi ^+_{AB}|]
\end{equation}
the spectrum of which is given by,  
\begin{equation}
\label{w2ev}
\frac{1}{3},\,\frac{1+x}{3},\,\frac{1-x}{3}.
\end{equation}
We also find that the corresponding  single qubit 
density matrix is the identity operator: 
$\hat\rho _A~=~\frac{1}{2}\,~(~\vert~\uparrow _A~\rangle~\langle~\uparrow_A~\vert~
+~\vert~\downarrow _A\rangle\langle \downarrow _A~\vert~)~=~\frac{1}{2}\,I_A$.  

There are two independent $q$-conditional entropies associated with three 
qubit states~\cite{Abe}, which are given by
\begin{eqnarray}
\label{AR2}
S^{(T)}_q(A|BC)&=&\frac{S^{(T)}_q(\hat\rho_{ABC})-S^{(T)}_q(\hat\rho_{BC})}
{1+(1-q)\,S^{(T)}_q(\hat\rho_{BC})}\nonumber \\
S^{(T)}_q(AB|C)&=&\frac{S^{(T)}_q(\hat\rho_{ABC})-S^{(T)}_q(\hat\rho_C)}{1+(1-q)\,
S^{(T)}_q(\hat\rho_C)}
\end{eqnarray}
and it was found~\cite{Abe}, in the case of three qubit 
Werner state,  that $S^{(T)}_q(A|BC)$ leads to stronger conditions
on separability than that identified from $S^{(T)}_q(AB|C)$. We examine here if this result holds 
good  for the three qubit symmetric states $\hat\rho _{ABC}$ also. 
The  $q$-conditional entropies $S^{(T)}_q(A|BC),\,S^{(T)}_q(AB|C)$ are evaluated
explicitly, by using Eqs.~(\ref{AR}), (\ref{w3ev}), (\ref{w2ev}) and (\ref{AR2}) 
\begin{eqnarray}
\label{ce2}
S^{(T)}_q(A|BC)&=&\frac{1}{q-1}\left[1-\frac{(\frac{1+3x}{4})^q+3\,(\frac{1-x}{4})^q}
{(\frac{1}{3})^q+(\frac{1+x}{3})^q+(\frac{1-x}{3})^q}\right]\nonumber\\
S^{(T)}_q(AB|C)&=&\frac{1}{q-1}\left[1-\frac{(\frac{1+3x}{4})^q+3\,(\frac{1-x}{4})^q}
{2\,(\frac{1}{2})^q}\right]
\end{eqnarray}

In Fig. 3(a) (3(b)) we plot the limiting values of the parameter $x$ for which 
$S^{(T)}_q(A|BC)=0\,(S^{(T)}_q(AB|C)=0)$ by varying $q$. From these implicit plots 
it is clearly seen that 
\begin{eqnarray*}
x\rightarrow \frac{1}{5}& {\rm as} &\lim_{q\rightarrow\infty}S^{(T)}_q(A|BC)=0\\
x\rightarrow \frac{1}{3}& {\rm as} &\lim_{q\rightarrow\infty}S^{(T)}_q(AB|C)=0.\\
\end{eqnarray*}
Thus the conditional entropy $S^{(T)}_q(A|BC)$ leads (in the limit 
$q\rightarrow\infty$) to 
\begin{equation}
\label{range}
0\leq x\leq\frac{1}{5}
\end{equation}
as the range of separability for the state $\hat\rho _{ABC}$ which is 
obviously stronger than the separability domain $0~\leq~x~\leq~\frac{1}{3}$ 
inferred by $S^{(T)}_q(AB|C)$.

\begin{figure}[h]
 \includegraphics*[width=2.15in,keepaspectratio]{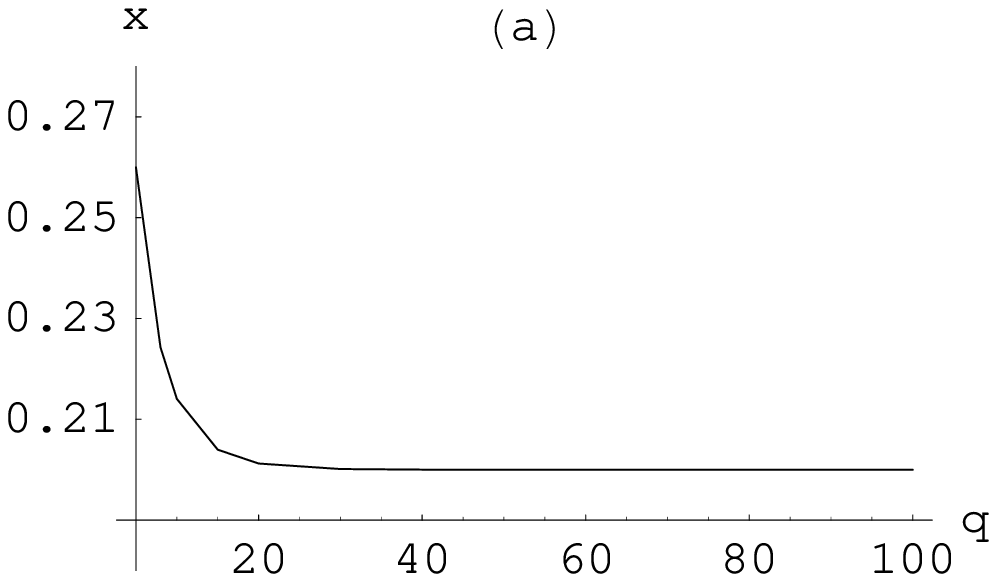}
\includegraphics*[width=2.15in,keepaspectratio]{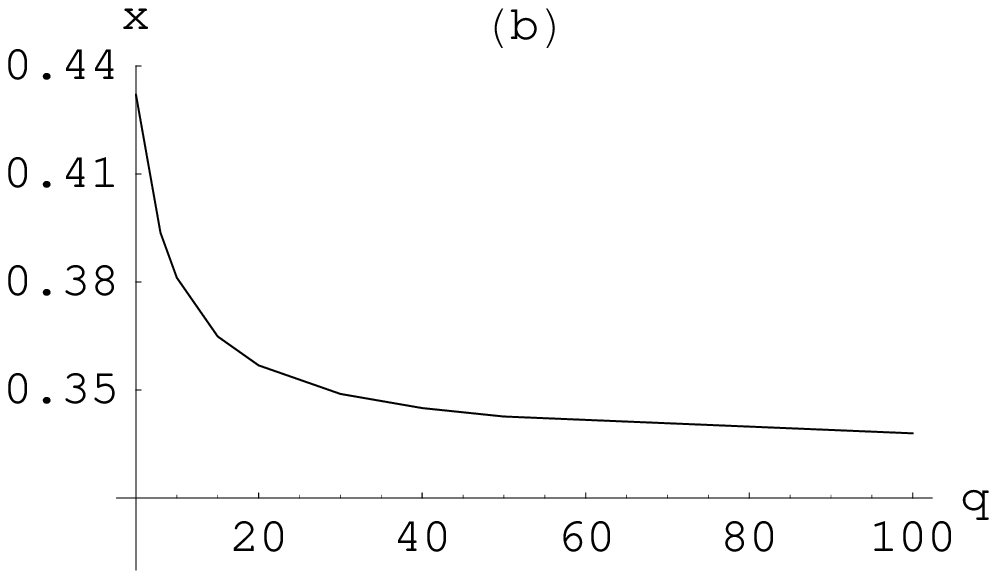}
 \caption{(a) Implicit plot of $S^{(T)}_q(A|BC)=0$ as a function of 
 $q$, corresponding to the three qubit state $\hat\rho_{ABC}$ 
 of Eq. (\ref{three}). In the limit $q\rightarrow\infty$ it 
 may be seen that $x\rightarrow \frac{1}{5}$. (b) Implicit 
 plot of $S^{(T)}_q(AB|C)=0$ as a function of  $q$. Here 
 $x\rightarrow \frac{1}{3}$ in the limit $q\rightarrow\infty$. 
 All quantities are dimensionless.}
 \label{fig:ce3}
\end{figure}

It may be noted that Eq.~(\ref{range}) obtained from $q$-conditional 
entropies serves only as a sufficient condition for separability of 
the state $\hat\rho _{ABC}$. The necessary and sufficient condition 
resulting from the Peres criterion: $0\leq x\leq 0.1547$ is obviously 
a strongest limitation on the separability.    

\section{ Separability of one parameter Symmetric multi-qubit W and GHZ states}
\label{sec:three}

In this section, we investigate quantum correlations in a $N$-particle generalization 
of the one parameter mixed states. 
\subsection{One parameter family of W states}
The symmetric one parameter mixed multiqubit states, involving a W-state, of 
our interest are 
\begin{equation}
\label{wstate3}
\hat\rho ^W_{A_1,A_2,A_3,\ldots,A_N}=\left(\frac{1-x}{N+1}\right)P_N+x\,
|W\rangle _N \,_N\langle W|
\end{equation}
and the corresponding reduced $N-n$ qubit density operator is given by, 
\begin{widetext} 
\begin{eqnarray}
\label{nredW}
\hat\rho ^W_{A_1,A_2,A_3,\ldots,A_{N-n}}&=&\left(\frac{1-x}{N-n+1}\right)P_{N-n}
 +\,\frac{x}{N}\left[(N-n)\left|W\right\rangle_{N-n} \,
_{N-n}\left\langle W\right|
+n\left|\downarrow_{A_1},\ldots,\downarrow_{A_{N-n}}\right\rangle
\left\langle \downarrow_{A_1},
\ldots,\downarrow_{A_{N-n}}\right|\right],\nonumber\\
& & \hskip 3.75in n~=~0,1,2,\ldots,N-2.
\end{eqnarray}
\end{widetext}
Here 
\begin{eqnarray*}
|W\rangle_N&=&|N/2,\,N/2-1\rangle=\frac{1}{\sqrt N}\,
[|\downarrow _{A_1},\uparrow _{A_2},
\ldots,\uparrow _{A_N}\rangle\\
& &+|\uparrow _{A_1},\downarrow _{A_2},\uparrow _{A_3},
\ldots,\uparrow _{A_N}\rangle+{\rm permutations}]
\end{eqnarray*}
and $P_N=\sum _{M=-N/2}^{N/2}\,|N/2,\,M\rangle\,\langle N/2,\,M |$.

From the discussions of Sec. II, it follows that the conditional entropy
\begin{widetext}
\begin{equation}
\label{conent}
S^{(T)}_q(A_1|A_2,A_3,\ldots,A_{N-n})=\frac{S^{(T)}_q(\hat\rho_{A_1,A_2,
A_3,\ldots , A_{N-n}})
-S^{(T)}_q(\hat\rho_{A_2,A_3,\ldots ,A_{N-n}})}{1+(1-q)\,S^{(T)}_q(\hat\rho_
{A_2,A_3,\ldots , A_{N-n}})};\ \ n=0,1,2,\ldots,N-2
\end{equation}
\end{widetext}
provides a stronger condition,  than that obtained from 
$S^{(T)}_q(A_1,\ldots,A_m|A_{m+1},\ldots,A_{N-n}),$ \break $m=2,3,\dots,N-n-1,$  
on the separability of multiparticle states 
$\hat\rho ^W_{A_1,A_2,A_3,\ldots,A_{N-n}}$.
In order to examine the asymptotic negativity of $S^{(T)}_q(A_1|A_2,A_3,\ldots,A_{N-n})$ 
we now proceed to evaluate the eigenvalues of $\hat\rho ^W_{A_1,A_2,A_3,\ldots,A_{N-n}}$ 
and its one qubit reduced marginal density matrix~\cite{footnote2} 
$\hat\rho ^W_{A_2,A_3,A_4,\ldots,A_{N-n}}
\equiv\hat\rho ^W_{A_1,A_2,A_3,\ldots,A_{N-n-1}}$.

The nonzero eigenvalues of $\hat\rho ^W_{A_1,A_2,A_3,\ldots,A_{N-n}}$are 
easily evaluated by expressing the state in the symmetric basis of total angular 
momentum $J=\frac{N-n}{2}:$
\begin{eqnarray}
& &\frac{1-x}{N-n+1}\,\,\,{((N-n-1)\,{\rm fold\,degenerate})},\nonumber\\ 
& &\frac{1-x}{N-n+1}+\frac{n\,x}{N},\,\,\, \frac{1-x}{N-n+1}+\frac{(N-n)x}{N}.
\end{eqnarray}

Using these eigenvalues the AR conditional entropy of Eq.~(\ref{conent}) 
is obtained as
\begin{widetext}
\begin{equation}
\label{conwstate}
S^{(T)}_q(A_1|A_2,A_3,\ldots,A_{N-n})=\frac{1}{q-1}\left[1-\frac{
(N-n-1)\left(\frac{1-x}{N-n+1}\right)^q+
\left(\frac{1-x}{N-n+1}+\frac{n\,x}{N}\right)^q+
\left(\frac{1-x}{N-n+1}+\frac{(N-n)x}{N}\right)^q}
{(N-n-2)\left(\frac{1-x}{N-n}\right)^q+\left(\frac{1-x}{N-n}
+\frac{(n+1)\,x}{N}\right)^q+
\left(\frac{1-x}{N-n}+\frac{(N-n-1)x}{N}\right)^q
}\right]
\end{equation}
\end{widetext}
The limiting value of the parameter $x$ satisfying 
$\lim_{q\rightarrow\infty}S^{(T)}_q(A_1|A_2,A_3,\ldots,A_{N-n})=0$ 
is determined by noting that only the maximum eigenvalue $p ^{\rm (max)}_{N-n}$ of 
$\hat\rho ^W_{A_1,A_2,A_3,\ldots,A_{N-n}}$ and $p^{\rm (max)}_{N-n-1}$ of 
$\hat\rho ^W_{A_2,A_3,\ldots,A_{N-n}}$ contribute in Eq.~(\ref{conwstate}) 
in the limit $q\rightarrow\infty$~\cite{Abe}. We thus find from the asymptotic negativity 
of $S^{(T)}_q(A_1|A_2,A_3,\ldots,A_{N-n})$ that the state $\hat\rho 
^W_{A_1,A_2,A_3,\ldots,A_{N-n}}$ 
is separable if 
\begin{equation}
\label{wineq}
0\leq x<\frac{N}{(N-n)^2+2N-n}.
\end{equation}
For $n=0,\, N=2$ we recover the separability condition $0\leq x\leq\frac{1}{4}$ 
for the two qubit state of Eq.(\ref{wstate2}) from the general 
result given in Eq. (\ref{wineq}).

\subsection{One parameter $N$-qubit  GHZ states}
\label{sec:}
We now proceed to find the $q$-entropic inference on the separability 
of  one parameter family  of symmetric states containing the maximally 
entangled  GHZ state:
\begin{equation}
\label{GHZstates}
\hat\rho^{{\rm GHZ}}_{A_1,A_2,A_3,\ldots,A_N}=\left(\frac{1-x}{N+1}\right)P_N
+x\,|{\rm GHZ}\rangle_N\,_N\langle {\rm GHZ}|
\end{equation}
where 
$$|{\rm GHZ}\rangle_N=\frac{1}{2}\,(|\uparrow _{A_1},\uparrow _{A_2},\ldots,\uparrow _{A_N}\rangle
+|\downarrow_{A_1},\downarrow _{A_2},\ldots,\downarrow _{A_N}\rangle ).$$
In order to determine $S^{(T)}_q(A_1|A_2,A_3,\ldots,A_{N})$ of the state 
$\hat\rho ^{{\rm GHZ}}_{A_1,A_2,A_3,\ldots,A_N}$ we now proceed to evaluate the 
eigenvalues of the given state and its one qubit reduced marginal density 
matrix, which is given by,
\begin{widetext}
\begin{equation}
\label{marGHZ}
\hat\rho ^{{\rm GHZ}}_{A_2,A_3,\ldots,A_N}=\left(\frac{1-x}{N}\right)P_{N-1}+
\frac{x}{2}\left[\left|\frac{N-1}{2},\,\frac{N-1}{2}\right\rangle
\left\langle\frac{N-1}{2},\,\,\frac{N-1}{2}\right|
+\left|\frac{N-1}{2},\,\,-\frac{N-1}{2}\right\rangle
\left\langle\frac{N-1}{2},\,\,-\frac{N-1}{2}\right|\right].
\end{equation}
\end{widetext}
The non-vanishing eigenvalues of the state $\hat\rho ^{{\rm GHZ}}_{A_1,A_2,A_3,\ldots,A_N}$ are 
evaluated as,
\begin{equation}
\frac{1-x}{N+1}\,\,(N \,{\rm fold\,degenerate}),\, \frac{1+Nx}{N+1}
\end{equation}
and that of its one qubit reduced subsystem $\hat\rho ^{{\rm GHZ}}_{A_1,A_2,A_3,\ldots,A_{N-1}}$ 
are, 
\begin{equation}
\frac{1-x}{N}\,\,((N-2)\,\, {\rm times}),\, \frac{2+x(N-2)}{2N}\,\,({\rm twice}).
\end{equation}
So, we obtain the $q$-conditional entropy associated with the $N$-qubit state 
$\hat\rho ^{{\rm GHZ}}_{A_1,A_2,A_3,\ldots,A_N}$ as,  
\begin{widetext}
\begin{equation}
S^{(T)}_q(A_1|A_2,A_3,\ldots,A_{N})=\frac{1}{q-1}\left[1-\frac{N\left(\frac{1-x}{N+1}\right)^q+
\left(\frac{1+Nx}{N+1}\right)^q}{(N-2)\left(\frac{1-x}{N}\right)^q+
2\left(\frac{2+x(N-2)}{2N}\right)^q}\right]
\end{equation}
\end{widetext}
As discussed earlier, we find the value of the parameter $x$, 
that marks the border of separability of the quantum state 
$\hat\rho ^{\rm GHZ}_{A_1,A_2,A_3,\ldots,A_N}$, by solving the equation 
$\lim_{q\rightarrow\infty}S^{(T)}_q(A_1|A_2,A_3,\ldots,A_{N})=0$, which leads to
$\left(1-\frac{p_{N}^{\rm (max)}}{p_{N-1}^{\rm (max)}}\right)=0$, in terms of the 
maximum eigenvalues $p_{N}^{\rm (max)}$, $p_{N-1}^{\rm (max)}$  of 
$\hat\rho ^{\rm GHZ}_{A_1,A_2,A_3,\ldots,A_N}$ and 
$\hat\rho ^{\rm GHZ}_{A_2,A_3,\ldots,A_{N-1}}$ respectively. So, we obtain 
the range of separability as 
\begin{equation}
\label{GHZrange}
0\leq x<\frac{2}{N^2+N+2}
\end{equation}
for the state $\hat\rho ^{\rm GHZ}_{A_1,A_2,A_3,\ldots,A_N}$.

Note that when $N=3$, the necessary and sufficient condition for separability 
$0\leq x\leq\frac{1}{7}$, 
inferred from the Peres criterion,  agrees with that obtained from the 
$q$-conditional entropy method (see Eq. (\ref {GHZrange})).
But for $N=4$, we obtain $0\leq x\leq 0.0625$ from the Peres criterion whereas Eq. (\ref 
{GHZrange}) 
gives $0\leq x\leq 0.0909$. In other words, the $q$-entropy result is weaker compared 
to that obtained from the Peres' criterion in one parameter family of  $N$ qubit GHZ 
states for $N>3.$
\section{Conclusion}
\label{sec:Conclusion}

We have employed the Abe-Rajagopal $q$-conditional entropy method
to characterize separability in one parameter family of symmetric $N$-qubit 
W and GHZ mixed states. It is identified that positivity of  
the $q$-conditional entropy,  $\displaystyle\lim_{q\rightarrow 
\infty}S^{(T)}_q(A_1|A_2,A_2\ldots),$ 
gives rise to strongest limitations on separability, 
 compared to that obtained from     
 the positivity of von Neumann conditional entropy ($q=1$ limit). 
In the case of symmetric one parameter family of 
two qubit states $\hat\rho_{AB}$, given by  Eq.~(\ref{wstate2}),   
and the three qubit state $\hat\rho ^{GHZ}_{A_1,A_2,A_3}$, we have recovered, using the 
$q$-entropy approach,  the necessary and sufficient conditions of separability  
on the parameter $x$. However, for the mixed W states, 
$\hat\rho ^{W}_{A_1,A_2,A_3,\ldots,A_{N-n}},\,n=0,1,\ldots,{N-2}$ 
and the GHZ  states $\hat\rho ^{GHZ}_{A_1,A_2,A_3,\ldots,A_N}$ with $N>3$, 
the range of separability, identified from the 
 $q$-conditional entropy approach,  is found to be 
 weaker compared to that obtained from the Peres' criterion. 
These general observations, concerning the separability of one-parameter family of symmetric 
states
are in confirmation with the results revealed by the 
detailed numerical investigations
performed on the full state space of arbitrary bipartite mixed states~\cite{Batle1, Batle2}.  
It would be illuminating to numerically investigate the 
$q$-conditional entropic behavior of more general 
multiqubit states. 
\begin{center}
{\bf ACKNOWLEDGEMENT}
\end{center}

\medskip

We gratefully acknowledge insightful discussions with Professor A. K. Rajagopal.


\begin{thebibliography}{1}
\bibitem{note} For a pure entangled state, the entropy of the whole system is zero: 
$S(\hat{\rho}_{AB})=0,$ and that of the subsystems (which are mixed) is non-zero: 
$S(\hat\rho_A)=S(\rho_B)>0.$
\bibitem{note2} The R\'{e}nyi entropy is  'additive', but not 'concave' for $q>1$, 
whereas the Tsallis entropy is concave for all values of $q>0$, however, 
it satisfies a modified additivity relation: 
$S^{(T)}_q(\hat\rho_A\otimes\hat\rho_B)=S^{(T)}_q(\hat\rho_A)+S^{(T)}_q(\hat\rho_B)
+~(1-q)~S^{(T)}_q(\hat\rho_A)S^{(T)}_q(\hat\rho_B),$ from which it is obvious that additivity 
follows only in the limit $q\rightarrow 1.$
\bibitem{note3} From Eq.~(\ref{AR}) it is clear that
$S^{(T)}_{q}(B|A)$ is positive whenever
${\rm Tr}\,[\hat\rho^q_{AB}]\leq {\rm Tr}\,[\hat\rho^q_{A}]$ for $q>1$ and  
 ${\rm Tr}\,[\hat\rho^q_{AB}]\geq{\rm Tr}\,[\hat\rho^q_{A}]$ for $0<q<1.$ 
Thus, concerning positivity of the conditional entropies, it may be seen that   
$S^{(T)}_{q}(B|A)\geq 0 \Longleftrightarrow S^{(R)}_{q}(B|A)\geq 0.$
\bibitem{footnote2} Subsystem density matrices of any randomly chosen $m~\leq~N$ 
  number of qubits, from a symmetric $N$ qubit state are all identical. More 
  specifically the two qubit marginal density matrices associated with the three 
  qubit symmetric state $\hat\rho _{ABC}$ satisfy: $\hat\rho _{AB}=
  \hat\rho _{BC}=\hat\rho _{AC}$.
\end{thebibliography}
\end{document}